\documentclass[sigconf]{acmart}

\setcopyright{acmlicensed}
\copyrightyear{2026}
\acmYear{2026}
\setcopyright{cc}
\setcctype{by}
\acmConference[MSR '26]{23rd International Conference on Mining Software Repositories}{April 13--14, 2026}{Rio de Janeiro, Brazil}
\acmBooktitle{23rd International Conference on Mining Software Repositories (MSR '26), April 13--14, 2026, Rio de Janeiro, Brazil}
\acmPrice{}
\acmDOI{10.1145/3793302.3793569}
\acmISBN{979-8-4007-2474-9/2026/04}

\usepackage{enumitem}

\begin{document}

\title{Reliability of AI Bots Footprints in GitHub Actions CI/CD Workflows}

\author{Syed Muhammad Ashhar Shah}
\email{25030009@lums.edu.pk}
\affiliation{
  \institution{Lahore University of Management Sciences}
  \city{Lahore}
  \country{Pakistan}
}

\author{Sehrish Habib}
\email{27100313@lums.edu.pk}
\affiliation{
  \institution{Lahore University of Management Sciences}
  \city{Lahore}
  \country{Pakistan}
}

\author{Muizz Hussain}
\email{26100368@lums.edu.pk}
\affiliation{
  \institution{Lahore University of Management Sciences}
  \city{Lahore}
  \country{Pakistan}
}

\author{Maryam Abdul Ghafoor}
\email{maryam.ghafoor@lums.edu.pk}
\affiliation{
  \institution{Lahore University of Management Sciences}
  \city{Lahore}
  \country{Pakistan}
}

\author{Abdul Ali Bangash}
\email{abdulali@lums.edu.pk}
\affiliation{
  \institution{Lahore University of Management Sciences}
  \city{Lahore}
  \country{Pakistan}
}

\begin{abstract}
Continuous Integration and Deployment (CI/CD) workflows are central to modern software delivery, yet the reliability of agentic AI bots operating within these workflows remain underexplored. Using pull requests (PRs), commits, and repositories from the AIDev dataset, we retrieved associated CI/CD workflow runs via the GitHub Actions API and analyzed 61,837 runs from 2,355 repositories, all triggered by PRs generated by five AI bots: Claude, Devin, Cursor, Copilot, and Codex. We observed substantial agent-dependent differences in workflow reliability, with Copilot and Codex achieving the highest success rates $\sim$93\% and $\sim$94\% respectively. At the repository level, we find a negative correlation between AI agent contribution frequency and workflow success rate, suggesting that a higher frequency of Agentic PRs may hinder CI/CD workflow reliability. We defined a taxonomy of 13 categories against 3,067 agentic PRs whose associated workflows failed, and observed a trend analysis that indicates visually observable shifts from functional to non-functional PR categories over time, although these trends are not statistically significant. Our findings motivate the need for actionable guidance on integrating AI agents into CI/CD workflows and prioritizing safeguards in workflows where failures are most likely to occur.
\end{abstract}

\maketitle

\keywords{Agentic AI, CI/CD, GitHub Actions, Workflow Analysis}

\section{Introduction}

Continuous Integration and Continuous Deployment (CI/CD) systems automate the build, test, and deployment stages of software workflows~\cite{hilton2016usage,vasilescu2015quality}. The rise of autonomous code-generating agents has ushered in \textit{Software Engineering 3.0 (SE3.0)}~\cite{qian2024sweagent}, where AI systems collaborate with developers across the entire software lifecycle. Tools such as Copilot, Codex, Claude, Cursor, and Devin use large language models (LLMs) to generate and modify code~\cite{chen2021evaluating,barke2023grounded}. Several agents now operate with near end-to-end autonomy, from interpreting software issues to submitting patches and deploying fixes~\cite{qian2024sweagent,yang2024devin}.
Despite rapid adoption, we still know little about how these AI agents behave inside CI/CD environments~\cite{2508.11867}. Existing work has studied LLMs for code generation, testing, debugging, and documentation~\cite{10.1145/3643991.3645084,arXiv:2507.10422}. However, the reliability of CI/CD workflows triggered by agentic pull requests, which we define in this study as the CI workflow success rate measured by the proportion of workflow runs that complete successfully, remains largely unexamined.

We address this knowledge gap through a large-scale empirical study of 61,837 GitHub Actions workflow executions triggered by 3,067 AI-agent PRs across open-source repositories. Our study is guided by two research questions: \\
\textbf{RQ1:} How does Agentic AI contributions affect CI/CD workflow outcomes? \\
\textbf{RQ2:} Among Agentic PRs that lead to workflow failures, which PR types are most prevalent, and how do they evolve over time? \\
 
Our work makes two primary contributions:
 (i) a quantitative analysis of workflow success and failure across AI agents and programming language types,
 (ii) a characterization and trend analysis of PRs that lead to CI/CD failures.
 To promote open-science, we have shared all our data, scripts, and experimental artifacts in a replication package on Zenodo~\cite{shah2025replication}.

\section{Background and Related Work}
\label{sec:background}

Recent work explored how AI integrates into software engineering tasks, with several studies analyzing its impact on code generation, debugging, automated testing, and documentation~\cite{10.1145/3643991.3645084,arXiv:2507.10422}. Mining-based analysis of GitHub repositories further examine how developers collaborate with AI-powered agents, revealing patterns in AI-assisted commits, PRs, and issue discussions~\cite{hilton2016usage,vasilescu2015quality}.
Foundational work on CI/CD reliability has examined why workflows fail. Prior work shows that CI/CD reliability depends heavily on CI configuration and usage. Gallaba et al.~\cite{gallaba2018use} found widespread misuse of Travis CI features and recurring anti-patterns that lead to fragile or misleading build outcomes, indicating that CI failures often stem from configuration rather than code defects. Beller et al.~\cite{beller2017oops} provides an early empirical study of Travis CI, showing how test failures propagate through automated builds. More recently, Aïdasso et al.~\cite{aïdasso2025illusionsuccessempiricalstudy} uncover “silent failures” and misleading success signals in industrial CI systems, highlighting the fragility of build reruns and reliability reporting.
Parallel work on AI-assisted software engineering identifies both productivity benefits and emerging risks. Cotroneo et al.~\cite{cotroneo2025humanwrittenvsaigeneratedcode} conduct a large-scale comparison of human-written and AI-generated code, showing that AI-produced artifacts exhibit different defects, vulnerabilities, and complexity profiles. Hassan et al.~\cite{hassan2025agenticsoftwareengineeringfoundational} outline the foundational principles of agentic software engineering and articulate open research challenges related to reliability, reproducibility, and safe autonomy.

In this study, we analyze 208,843 CI/CD workflow runs triggered by 33,596 PRs generated by five agentic AI bots, spanning 88,576 commits across 2,355 GitHub OSS repositories. From these, we focus on 61,837 workflows where the AI agent triggered the CI/CD workflow through a PR and identify 3,067 PRs that result in workflow failures. We use topic modeling to develop a taxonomy of 13 PR types linked to failed workflows and examine how the percentage composition of each PR type evolve over time.

\begin{figure*}[t]
    \centering
    \includegraphics[width=.6\textwidth]{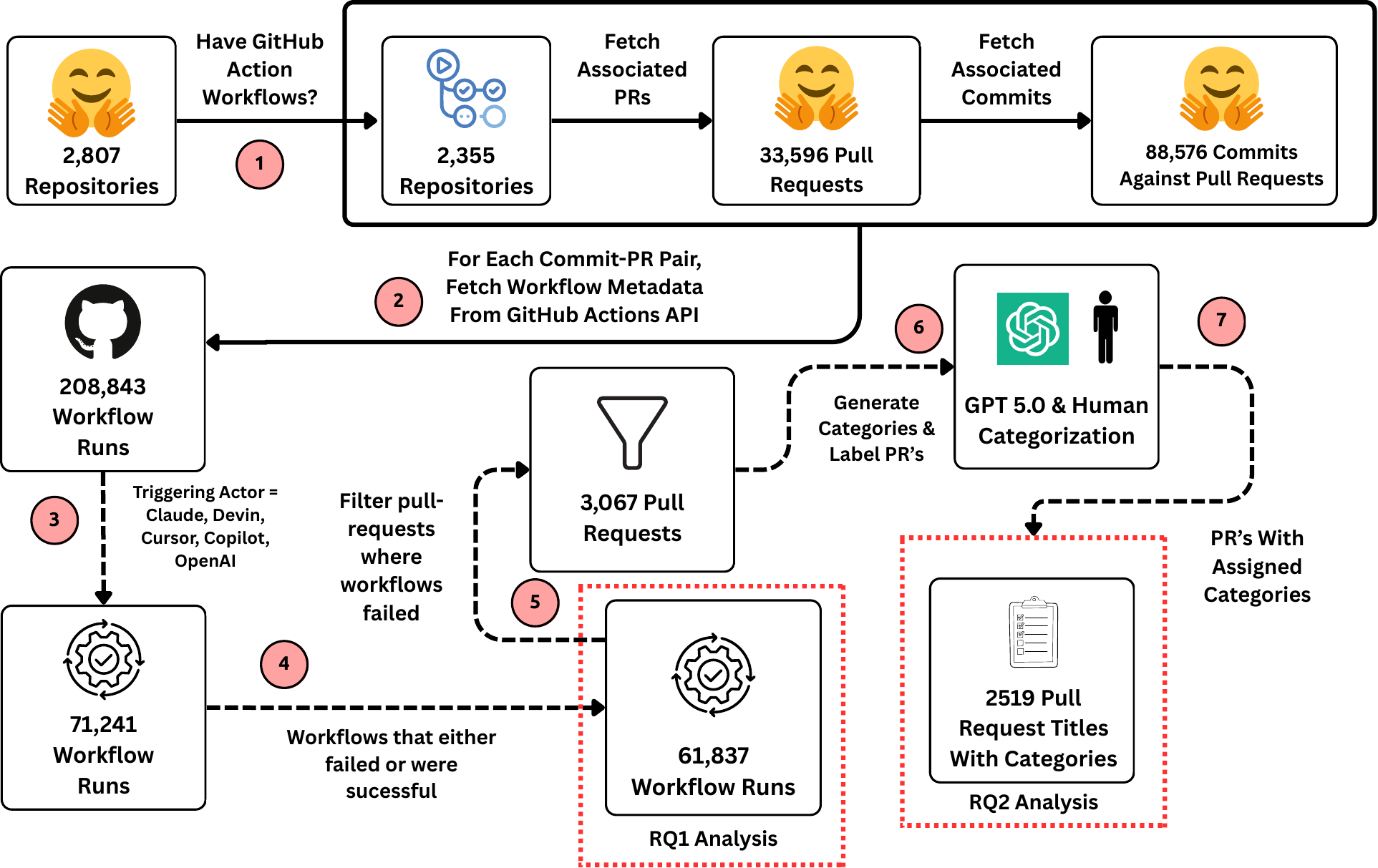}
    \caption{Overview of the CI/CD Reliability Analysis Workflow}
    \label{fig:methodology_diagram}
\end{figure*}

\section{Methodology}
\label{sec:methodology}

\subsection{Data Collection} 

We use the AIDev dataset~\cite{aidev_dataset}, a large-scale collection of GitHub repositories curated for studying AI-driven software development in OSS GitHub repositories. The dataset provides metadata such as commits, pull requests (PRs), and issues, enabling fine-grained analysis of agentic AI bots. We load the dataset into PostgreSQL for SQL-based retrieval and transformation.

We focus on the 2,807 repositories listed in the \textit{pr\_repositories} table of the AIDev dataset, which explicitly excludes toy, inactive, or artificially generated projects. To retrieve the relevant GitHub Actions workflow files, we search for YAML files (\textit{.yml}) within each repository's \textit{.github/workflows} directory. After this filtering step, 2,355 repositories remain for further analysis. Furthermore, from the \textit{pull\_request} and \textit{pr\_commits} tables of the AIDev dataset, we extract 33,596 pull requests and 88,576 commits corresponding to repositories that contain at least one valid CI/CD workflow.

Each PR has multiple commits, we distribute these commits into commit-PR pairs. For each commit–PR pair, we collect workflow execution metadata, including trigger type, which denotes the GitHub event that initiated the workflow, job status, which indicates the final execution outcome (e.g., success or failure), and duration, which measures the total time taken to complete the workflow, via the GitHub Actions API, resulting in 208,843 workflow runs. To complement the AIDev dataset, we filter these runs to those initiated by the PRs made by the five agentic AI bots (Claude, Devin, Cursor, Copilot, and Codex), yielding 71,241 workflow runs. We further restrict our analysis to workflows that have either succeeded or failed, producing a final set of 61,837 workflow runs that we use in our quantitative and qualitative analysis.

Figure~\ref{fig:methodology_diagram} presents the end-to-end pipeline of our methodology, i.e., from data extraction and repository filtering to workflow metadata retrieval and analysis.

\subsection{Language type tagging}

To understand the success of Agentic PRs in the CI/CD workflows, we categorize the PRs into their repository's language type (high-level language or low-level language).
To figure out the language of a repository, we use the pr\_repository table of the AIDev dataset which has this information.
Following the standard convention~\cite{aho2006compilers}, we group languages into \emph{low-level} (C, C++, C\#, Rust, Zig, Assembly) and \emph{high-level} (Python, Java, JavaScript, TypeScript, Ruby, PHP, Go, Kotlin, Swift, R, Dart, HTML, CSS, Shell, and others).
After this categorization, we tag 61,837 workflows with a language type (i.e., the language used in the repository of their PR), we find that 53,087 workflows use high-level languages, 8,192 workflows use low-level languages, and the remaining 558 had no language specified.

\subsection{Workflow success rate}
For each repository, we compute a workflow success rate by dividing the number of successful workflow runs by the number of total workflow runs. We then examine how the workflow success rate varies with the frequency of agentic PRs, enabling us to assess whether higher levels of AI contribution are associated with CI/CD workflow success/failure rates.

\subsection{PRs Categorization}

We categorize the PRs that trigger failing CI/CD workflows. In our dataset, \textbf{3,067 PRs} trigger CI/CD workflows that fail to resolve.

To categorize PRs, we sample 548 PRs from 3,067 PRs (with 99\% confidence level and a 5\% margin of error) and perform initial semantic clustering on the PR titles using \textit{GPT~5.0}~\cite{Zhao_2024_Asset}. Two authors then refine these clusters through manual sorting, resolving disagreements via a negotiated agreement process~\cite{Braun_Agreement}. This process produced final categories of 13 PR types: \textit{Bug Fixes}, \textit{Testing \& Quality Assurance}, \textit{New Features \& Enhancements}, \textit{APIs, SDKs \& Integrations}, \textit{User Interface \& User Experience}, \textit{Configuration \& Infrastructure}, \textit{Refactoring \& Code Quality}, \textit{Documentation \& Examples}, \textit{Security \& Authentication}, \textit{Performance \& Optimization}, \textit{Regular Maintenance \& Miscellaneous}, \textit{Tools, Utilities \& CLI}, and \textit{Continuous Integration \& Continuous Deployment}.

To perform close-card sorting, we then provide \textbf{GPT~5.0} the remaining PRs' titles and the finalized 13 PR categories from the previous step. The model then assigned each PR to the most relevant category. 
To further validate the labeling of PRs, we manually sample \textbf{334 PRs} out of \textbf{2,519 PRs} (with 95\% confidence level, 5\% margin of error). Two authors independently coded the samples, and we found a strong inter-rater agreement between human and GPT~5.0, through a \textbf{Cohen’s Kappa of 0.88}. This step resulted in a labeled dataset of \textbf{3,067 PRs} with assigned PR categories.

\begin{table}[H]
    \centering
    \caption{Agent Success Rate Distribution by Language Level}
    
    \label{tab:agent_performance}
    \resizebox{1\columnwidth}{!}{%
    \begin{tabular}{|l|r|r|r|r|r|r|} 
        \hline
        \textbf{Agent} & \multicolumn{2}{|c|}{\textbf{High-Level}} & \multicolumn{2}{|c|}{\textbf{Low-Level}} & \multicolumn{2}{|c|}{\textbf{Total}} \\
        \cline{2-7}
        & \textbf{Success Rate} & \textbf{Total Runs} & \textbf{Success Rate} & \textbf{Total Runs} & \textbf{Success Rate} & \textbf{Total Runs} \\
        \hline
        \textbf{Claude} & 63.89\% & 36.00 & 100.00\% & 1.00 & 64.86\% & 37.00 \\
        \textbf{Copilot} & 93.10\% & 9,349.00 & 93.62\% & 4,830.00 & 93.28\% & 14,179.00 \\
        \textbf{Cursor} & 73.29\% & 2,696.00 & 65.07\% & 335.00 & 72.39\% & 3,031.00 \\
        \textbf{Devin} & 78.03\% & 40,921.00 & 69.05\% & 2,931.00 & 77.43\% & 43,852.00 \\
        \textbf{Codex} & 92.94\% & 85.00 & 95.79\% & 95.00 & 94.44\% & 180.00 \\
        \hline
    \end{tabular}%
    }
\end{table}

\section{Results}
\label{sec:results}

\subsection{CI/CD Reliability Under Increasing AI Contributions (RQ1)}

\subsubsection{Cross Agent and Language Type Reliability}

\textbf{In terms of workflow success rate, there is a significant difference between the agents, both for high-level language contributions and low-level language contributions.}
We find that Copilot and Codex achieves the highest success at 93.28\% and 94.44\%, while Claude had the lowest reliability at 64.86\%. Devin exhibits moderate reliability at 77.43\%, despite contributing most (71.56\%) of the workflows runs.
We detail these stats in Table~\ref{tab:agent_performance}, 

A chi-square test indicates statistically significant differences in workflow success rates across all agents ($p < 0.01$, $Cramer's~V = 0.177$). Pairwise post-hoc Fisher’s exact tests with Benjamini--Hochberg correction further confirm these differences. Copilot has significantly higher odds of workflow success than Claude ($\text{OR}=7.53$), Cursor ($\text{OR}=5.29$), and Devin ($\text{OR}=4.05$), while showing no statistically significant difference relative to Codex after correction ($\text{OR}=0.816$). Here, an Odds Ratio (OR) greater than 1 indicates that the first agent has higher odds of success than the second, while OR less than 1 indicates lower odds.

\textbf{Our analysis in the context of programming language types show that agents have a slightly higher CI/CD reliability success on PRs for low-level languages (86.7\%) compared to high-level languages (82.4\%)}. 
A chi-square and effect size test confirms that this difference is statistically significant with a small effect size ($p < 0.01$, $Cramer's~V = 0.028$). The small effect size shows programming language might not be the major factor for cross agent differences.

\subsubsection{Relation of Agent PR Frequency with Workflow Outcomes}

\begin{figure}[t]
\centering
\includegraphics[width=0.9\columnwidth]{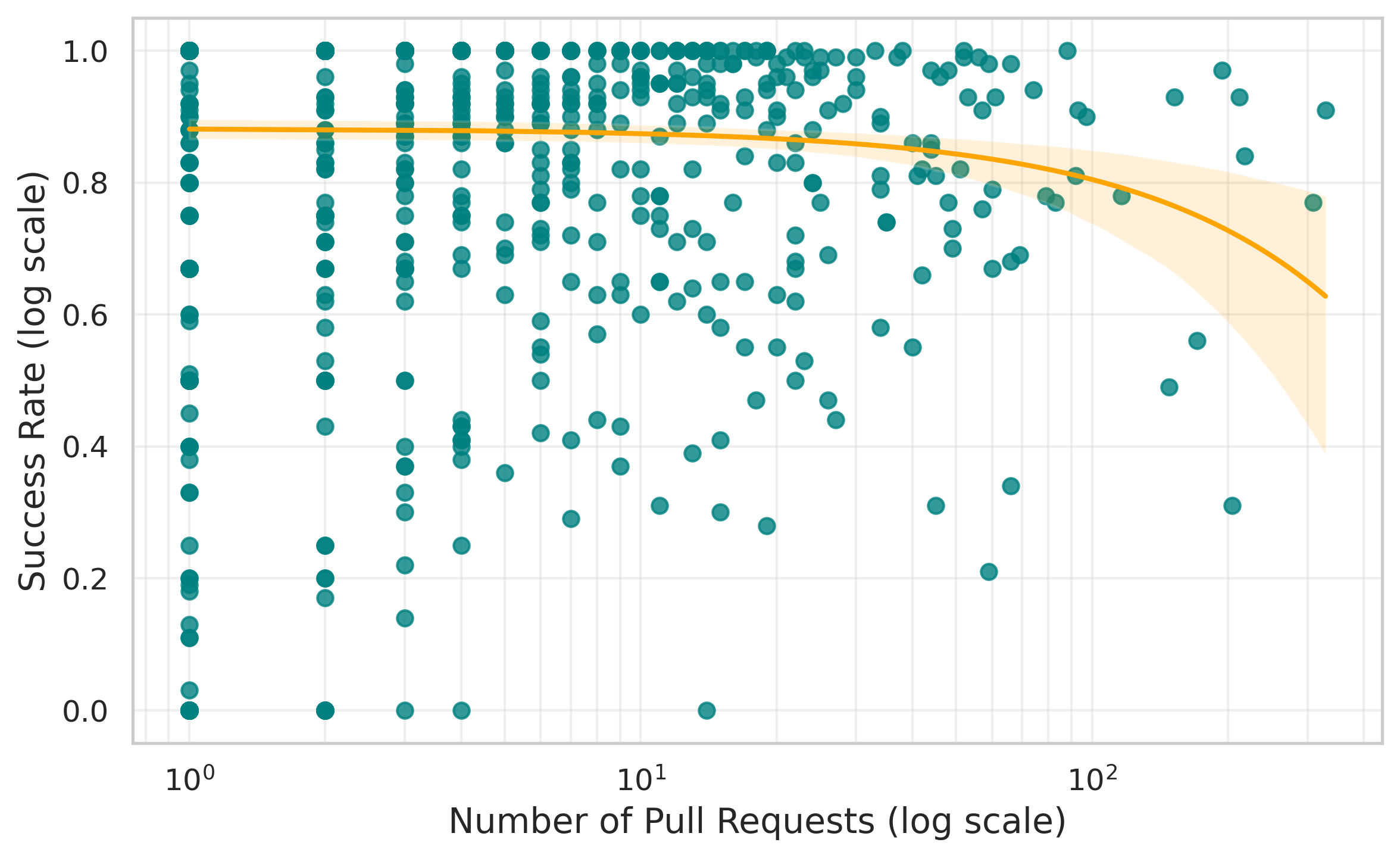}

\caption{Scatter plot of Agentic PRs vs. workflow success rate per repository}
\label{fig:rq1_contribution_intensity}
\end{figure}

We examine the relationship between AI agents contribution frequency and CI/CD workflow success rate at the repository level. Figure~\ref{fig:rq1_contribution_intensity} presents a logarithmic scatter plot of the number of Agentic PRs per repository against the corresponding workflow success rate.

The results show a weak negative corelation between Agent PR frequency and workflow success rate (Spearman $\rho = -0.34$, $p < 0.01$), indicating that \textbf{repositories with more Agentic PRs tend to exhibit lower workflow success rates}. This suggests that increased agentic activity \textbf{may} degrade CI/CD performance.

\subsection{Temporal Trends of PR Categories (RQ2)}

In RQ2, we analyze the PR category-wise differences and inspect, through a trend analysis, how they evolve over-time.
Specifically, we look at \textbf{2,519 PRs that triggered 9,012 failed workflows}.

\paragraph{Failure Category Distribution} 
Workflow failures are unevenly distributed across PR types, with higher proportions observed for \textit{Bug Fixes} (17.57\%), \textit{UI/UX} (11.64\%), \textit{New Features \& Enhancements} (10.26\%), \textit{Refactoring} (10.04\%), and \textit{Configuration \& Infrastructure} (8.04\%). In contrast, categories such as \textit{Security \& Authentication} (3.59\%), \textit{CI/CD} (3.42\%), and \textit{Tools \& CLI} (1.95\%) account for substantially smaller shares. A chi-square goodness-of-fit test confirms that workflow failures are unevenly distributed across PR categories ($p < 0.01$), indicating failures are not uniformly spread across different contribution types.

\paragraph{Temporal Trends in Failure Proportions}

\begin{figure}[t]
    \centering
    \includegraphics[width=0.9\columnwidth]{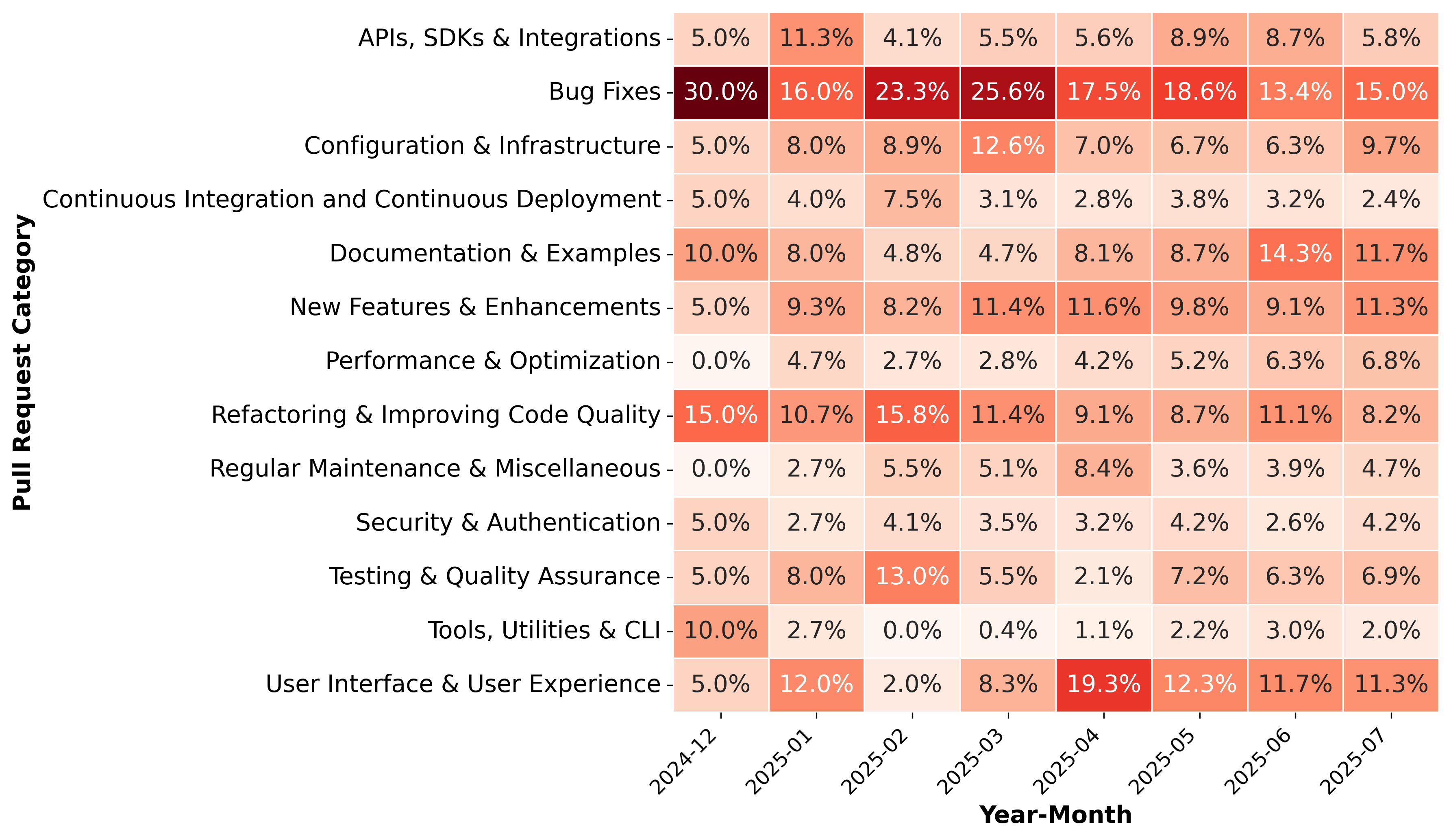}
    
    \caption{Month-by-Month Trends in PR Categories Leading to Workflow Failures (Dec 2024 –Jul 2025)}
    \label{fig:rq2_trends}
\end{figure}

The heatmap in Figure~\ref{fig:rq2_trends} shows evolving patterns in the PR contribution types that trigger workflow failures, with each cell representing the \% of PRs for that month. Functional categories, such as \textit{Bug Fixes} and \textit{New Features \& Enhancements}, consistently account for a large share of failures, with recurring peaks (e.g., Bug Fixes in Dec 2024 and Mar 2025).

Non-functional categories, such as \textit{Documentation \& Examples} and \textit{UI/UX}, show visually increasing patterns from May to July 2025. However, Mann--Kendall trend tests on the PR categories (with Benjamini--Hochberg correction) reveals no statistically significant monotonic trends across 8 months. While \textit{Performance \& Optimization} shows the strongest positive signal (highest Kendall’s $\tau$), its still not significant, indicating emerging but unconfirmed shifts in CI/CD failures.

\section{Threats to Validity}
\label{sec:threats}

\textit{Internal Validity.} Our analysis uses PR titles to classify the types of PRs associated with workflow failures. PR categories were generated using GPT-5.0, which may sometimes misclassify PRs. To mitigate this risk, we conducted manual human validation and observed strong inter-rater agreement between human annotations and the AI model (Cohen’s $\kappa = 0.88$, indicating high consistency in PR categorization).

\textit{External Validity.} Our dataset consists mainly of public GitHub repositories. Consequently, the findings might not fully generalize to private or enterprise CI/CD environments, which typically involve more complex workflows and processes.

\textit{Construct Validity.} We attribute agent activity based on commit authorship metadata; however, hybrid human–AI edits within the same PR might introduce minor inaccuracies. Additionally, we assign each PR to a single category for analysis, even though some PRs could span multiple types of changes.

\textit{Conclusion Validity.} Our statistical analysis assume that workflow runs are independent. In practice, repositories might contain multiple related workflows, introducing potential dependencies. Future work could employ hierarchical or mixed-effects models to account for these nested executions more accurately.

\section{Implications}
\label{sec:implications}

\textit{For software developers and maintainers}, the success rate of CI/CD workflows triggered by Agentic PRs varies across agents. Copilot and Codex generally perform well, whereas Cursor and Claude cause failures more frequently. Developers can use these insights to implement AI-aware validation gates and automated review processes to improve CI/CD workflow reliability.

\textit{For CI/CD engineers}, detecting and managing agentic AI commits separately can enable early anomaly detection. Our findings can help engineers prioritize monitoring PR types that more frequently trigger workflow failures and design CI/CD configurations to handle AI-generated code failures more effectively.

\textit{For AI model developers and researchers}, our findings identify the types of PRs that most frequently lead to workflow failures, pinpointing areas where AI agents face challenges in CI/CD workflows. By leveraging these insights, to fine-tune models on CI/CD-specific data, developers can enhance model reliability and foster safer, more robust AI coding assistants~\cite{10.1145/3657054.3664243, nahar2023authorship, 11029412}.

\section{Ethical Considerations}
We used the AIDev dataset and cited its authors throughout the paper to give appropriate credit. 

\section{Conclusion}
\label{sec:conclusion}

This study provides a large-scale empirical analysis of 61,837 GitHub Actions workflow runs to evaluate the reliability of agentic AI bots' contributions in CI/CD workflows. Our findings reveal that reliability is primarily agent-dependent: while Copilot and Codex achieve success rates exceeding 93\%, other agents like Claude and Cursor show a significantly higher tendency for failure. We also observe a weak negative correlation between agentic contribution frequency and workflow success rate, suggesting that increased agentic activity may lead to more failures in CI/CD workflows.

Our analysis of 3,067 PRs by their categories shows that agentic PRs are causing frequent CI/CD failures in functional categories, such as \textit{Bug Fixes}. Over time, however, failures in non-functional categories, such as \textit{UI/UX} and \textit{Performance \& Optimization}, also exhibit a growing trend, even though these increases are not statistically significant. Our results highlight the need for AI-aware validation gates and targeted monitoring of both functional and non-functional AI-based code changes. Future work will investigate the causes of agent-specific performance gaps and extend this analysis to private enterprise code bases to advance the understanding of Software Engineering 3.0.

\bibliographystyle{ACM-Reference-Format}
\bibliography{references}

\end{document}